\documentclass[12pt,preprint]{aastex}

\usepackage{emulateapj5}

\newcommand{\sbs}{SBS\,0335$-$052}
\newcommand{\simgt}{\lower.5ex\hbox{$\;\buildrel>\over\sim\;$}}
\newcommand{\simlt}{\lower.5ex\hbox{$\;\buildrel<\over\sim\;$}}

\slugcomment{}

\shorttitle{Templates for submm sources}
\shortauthors{Hunt \& Maiolino}

\begin{document}


\title{Improved Templates for Photometric Redshifts\\
    of Submm Sources}


\author{L. K. Hunt\altaffilmark{1}} \and \author{R. Maiolino\altaffilmark{2}}

\altaffiltext{1}{INAF--Istituto di Radioastronomia, Largo E. Fermi 5, 50125 Firenze, Italy;
{\tt hunt@arcetri.astro.it}}
\altaffiltext{2}{INAF--Osservatorio Astrofisico di Arcetri, Largo E. Fermi 5, 50125 Firenze, Italy;
{\tt maiolino@arcetri.astro.it}}



\begin{abstract}
There is growing evidence that some star-forming galaxies
at $z>1$ are characterized by high efficiencies and 
specific star formation rates.
In the local universe,
these traits are shared by ``active'' Blue Compact Dwarf galaxies (BCDs)
with compact and dense star-forming regions. The 
Spectral Energy Distributions (SEDs) of these BCDs are dominated 
by young massive star clusters, embedded in a cocoon of dust.
In this Letter, we incorporate these BCD SEDs as templates for
two samples of high-redshift galaxy populations selected at submm wavelengths.
Because of the severe absorption of the optical
light, the featureless mid-infrared spectrum,
and the relatively flat radio continuum, the dusty star-cluster SEDs 
are good approximations to most of the submm sources in our samples.
In most cases, the active BCD SEDs fit the observed photometric points 
better than the ``standard'' templates, M\,82 and Arp\,220,
and predict
photometric redshifts significantly closer to the spectroscopic ones.
Our results strongly suggest that the embedded dusty star clusters
in BCD galaxies are superior to other local templates such as M\,82 and Arp\,220
in fitting distant submm starburst galaxies. 
\end{abstract}





\section{Introduction}

The extension of the photometric redshift technique to the far-infrared (FIR) and
radio spectral ranges has proven to be a useful tool to investigate
the population of distant sub-millimeter (submm) sources 
\citep{smail97,barger98,hughes98} which, 
because of heavy dust obscuration, may have optical counterparts
that are too faint for a spectroscopic identification
\citep[e.g.][]{barger99}.
This technique, first proposed by \citet{carilli99}, initially exploited
only the radio--to--submm spectral index as a redshift indicator,
and is still hotly debated because of the
degeneracy between increasing redshift and decreasing dust temperature 
\citep{blain99,blain03,aretxaga05}.
Nevertheless, it has recently been
extended also to other bands \citep[][]{greve04,eales03},
including new Spitzer data \citep[e.g.][]{appleton04,frayer04}.
The templates commonly used to fit the FIR--to--radio SED of high
redshift dusty galaxies are those of well studied, evolved starburst galaxies
in the local universe (typically Arp220 and M82), or Active Galactic Nuclei (AGNs)
\citep[e.g.][]{aretxaga03}. This approach assumes that
dusty starburst galaxies at high redshift are characterized by the
same physical properties as local ``standard'' starburst galaxies. 

However, there is growing evidence that some 
starburst galaxies at z$>$1 are characterized by a different regime
of star formation.
\cite{bauer05} have shown that the specific star formation rate
(SSFR$\equiv$SFR/galaxy mass) at z$>$1 is significantly higher than observed
in the local universe \citep[see also][]{feulner05,greve05}. 
New Spitzer data reveal that
a significant fraction of high-redshift sources 
are characterized by the absence of PAH features in their
mid-IR spectra \citep{houck05}. 
\cite{houck05} interpret these observations as a
possible indication for AGN-powered galaxies, but a lack of PAH features 
is also observed in extragalactic extremely compact
star forming regions \citep[][]{thuan99,houck04,galliano05}. 
Finally, 
many high-z submm galaxies show dust obscuration much more extreme than
observed in local ``standard'' starburst templates \citep{hammer}, and
may more closely resemble highly efficient
maximum-intensity starbursts \citep{meurer97,somerville01,greve05}.

The Blue Compact Dwarf (BCD) galaxies found in the local universe
are certainly less luminous and much less massive than the submm
starburst galaxies found at high redshift.
They are also more metal poor, and 
therefore distinct from the massive ``SCUBA galaxies'' and Lyman-break
systems which are already chemically enriched at $z\la3$ 
\citep{pettini01,tecza,shapley04}.
However, a subclass of them, ``active'' BCDs \citep{hirashita04},
seems to share one characteristic with the high-redshift
submm population: extreme star forming modes.
Active BCDs host extremely compact, very young, super star
clusters, which are responsible for a very efficient SSFR,
and which dominate
their Spectral Energy Distributions (SEDs) \citep{vanzi04,hunt05}. 
The intense SSFR and
the compactness of the star forming regions makes the infrared SEDs
of active BCDs quite different from
``classical'' starburst galaxies, often being broader and/or warmer, 
and featureless
\citep[][]{hunt05}. The youth of these systems tends to make
the radio emission predominantly thermal rather than synchrotron
\citep{klein91,hunt05}. Super star clusters in BCDs can also be heavily embedded
in dust, which makes the stellar and nebular radiation heavily
absorbed even in the near-IR \citep[][]{hunt01}. These extreme properties
of BCDs suggest that they may represent a scaled down version
of the extreme starburst galaxies at high redshift. In particular,
BCDs could provide better templates for the SEDs of distant
starburst galaxies, thus enabling more accurate photometric redshifts.

In this Letter we test this hypothesis by comparing the SEDs
of local BCDs \citep[][]{hunt05} with the photometric measurements
of high redshift submm galaxies having secure spectroscopic identifications
and redshifts.
We show that BCD templates
generally provide a better match to the SEDs of submm/SCUBA galaxies and improve
significantly the accuracy of photometric redshifts. 

\section{The BCD SED Templates \label{sec:seds}}

The SEDs from 5\,$\mu$m to 20\,cm 
of low-metallicity ``active'' BCDs are well modelled by
dusty massive star clusters
\citep{hunt05}.
We have taken three cases from \cite{hunt05}: NGC\,5253, II\,Zw\,40, 
and SBS0335$-$052, which 
are characterized by different amounts of
dust extinction: $A_V$\,=\,10, 20, and 30\,mag, respectively.
Because the SED of SBS0335$-$052 is very unusual, peaking at $\sim$30\,$\mu$m,
we have also included another best-fit model for II\,Zw\,40 with $A_V\,=\,$30\,mag.
Also for II\,Zw\,40, we adopted two possible values
for the radio slope: pure thermal $\alpha=-0.1$,  and supernova-like
$\alpha=-0.5$, both of which are consistent with the data. 

Rather than the entire observed BCD SED,
we adopt here only the dusty star cluster SED, as given by \citet{hunt05},
which are based on {\it DUSTY} \citep{elitzur} models.
Because of the large optical depths,
the rest-wavelength optical bands in these SED templates are highly
reddened and absorbed.
The observed optical SEDs of BCDs are not well modelled by 
the dusty clusters, because
the optical light arises from regions outside of the clusters, or
at least from regions which are not subject to their high extinction. 
We do not consider such emission in our SED templates.

The BCDs which we are using as SEDs are generally metal poor.
However,
the ISM metallicity determines, at least in part, {\it only} the dust content
\citep[e.g.,][]{edmunds}, so
that it becomes a scaling factor, rather than a shaping factor in the SED.
Observationally, these star-cluster SEDs are 
applicable to a wide range of metal abundances, from highly sub-solar
(\sbs, $\sim$ 2.5\% $Z_\odot$) to $\sim$solar (He\,2$-$10)
\citep{hunt05}.
Hence, the SEDs we are using to model high-z sources are 
essentially independent of metallicity. 

\section{BCDs versus Submm Galaxies with Spitzer Data \label{sec:spitzer}}

\begin{figure}
\hbox{\includegraphics[angle=0,width=0.385\linewidth,bb=20 145 590 645]{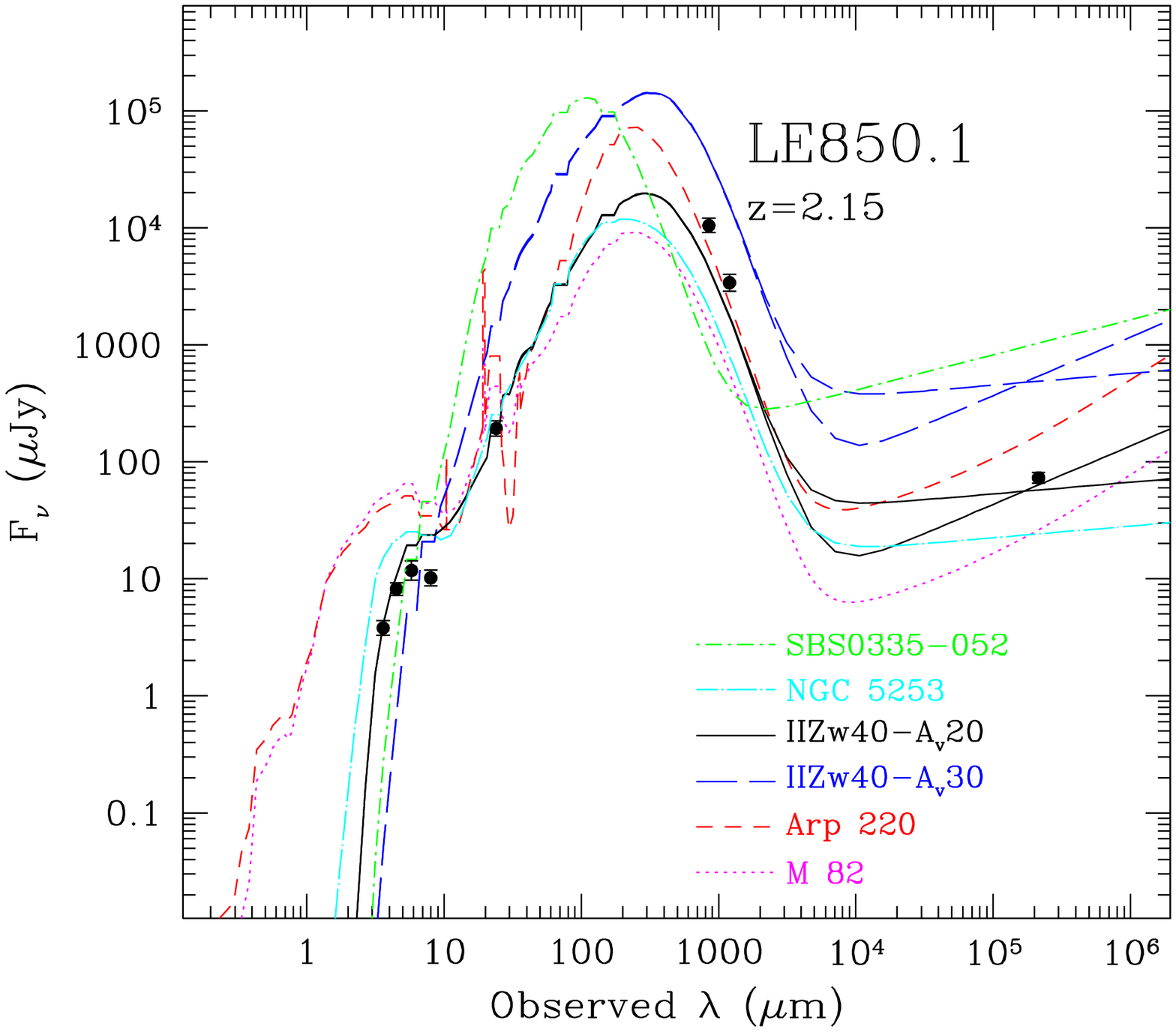}
\hspace{-1.45cm} 
\includegraphics[angle=0,width=0.385\linewidth,bb=20 145 590 645]{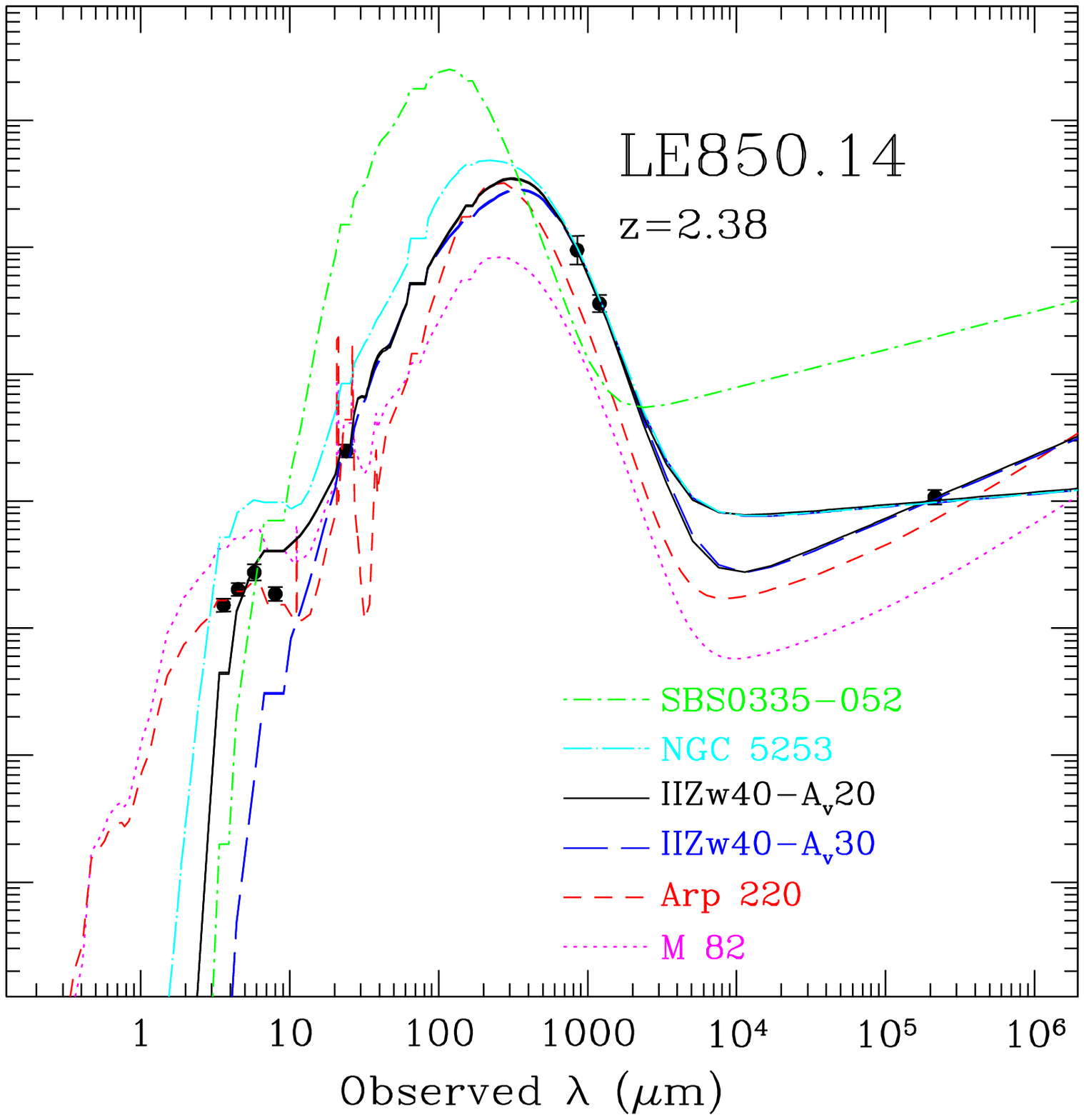} 
\hspace{-1.45cm}
\includegraphics[angle=0,width=0.385\linewidth,bb=20 145 590 645]{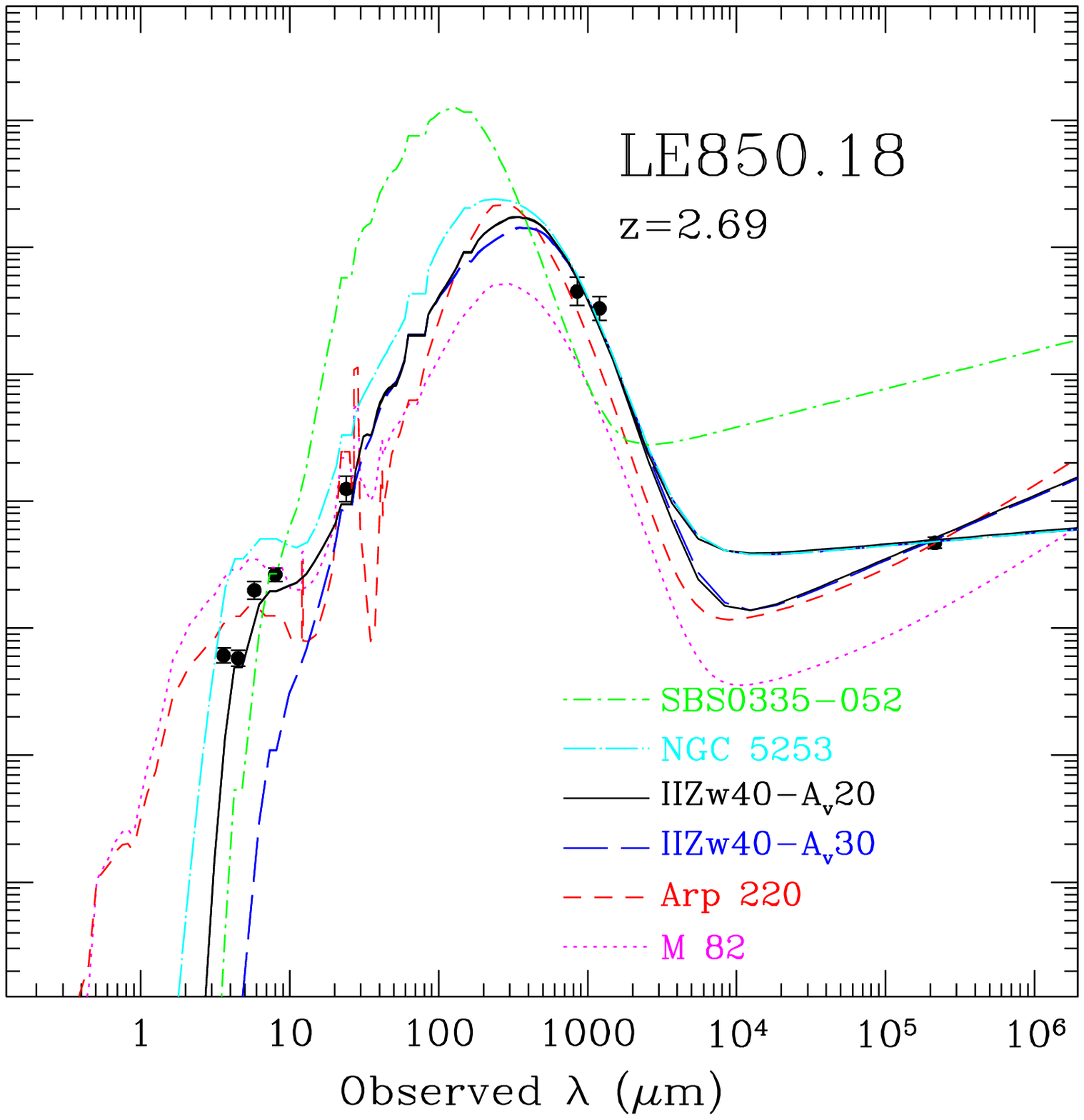} }
\caption{SEDs of radio-identified submm sources \citep{scott02,ivison02,egami04,ivison04},
with various starburst templates normalized to the data (see text).
The best-fit BCD SED templates are II\,Zw\,40 ($A_V=$20\,mag, $\alpha=-0.5$) for
LE\,850.1 and II\,Zw\,40 ($A_V=$20\,mag, $\alpha=-0.1$) for LE\,850.8 and LE\,850.14.
\label{fig:seds}}
\end{figure}

A few submm sources \citep[e.g.,][]{smail97,barger98,hughes98} 
have recently been observed with Spitzer
\citep[][]{egami04,frayer04,ivison04,higdon04}. Such new data
allow a more detailed analysis of the IR--to--radio SED
of these distant starburst galaxies than previously possible. Among the submm galaxies
observed with Spitzer we have selected those with secure
spectroscopic redshifts\footnote{A seventh source would
also be available, but it is lensed and therefore more complex
to interpret.}. These six sources and their basic
properties are listed in Table \ref{tab:zphot}. None of them
seem to host powerful AGN, at least according to their optical
spectra. Due to the limited space available in this Letter, 
the photometric points for only three of the six sources
are plotted in Fig. \ref{fig:seds}. 

We have fit the observed SEDs with the six templates
described in $\S$\ref{sec:seds}.
The ``fit'' is a scaling normalization to minimize residuals, and 
a calculation of the remaining (logarithmic) rms deviations.
For comparison with the templates
adopted in the past, we also fit the SEDs with the spectrophotometric
models of M\,82 and Arp\,220 given by \cite{bressan02}.
In total, we consider eight templates: six variations of BCD star clusters,
M\,82, and Arp\,220.
In Fig. \ref{fig:seds} we show 
examples of the various templates redshifted to the spectroscopic
redshift and scaled to best fit the photometric measurements.
In Table \ref{tab:zphot} we give the rms residuals for the best-fit template
evaluated at the spectroscopic redshift.
In four of the six cases, the BCD templates fit significantly better
the observed photometric points than either M\,82 or Arp\,220.

We have also tested the capability of each template to reproduce 
the spectroscopic redshift through the photometric redshift technique.
First, the SEDs were convolved with the Spitzer filters.
Then, by finding the
minimum of $\chi ^2$ residuals as a function of redshift for the
various templates, we derived the photometric redshift for each source. 
Table \ref{tab:zphot} reports the photometric redshifts obtained for each template.
It is impressive that the BCD templates {\it always} provide a
photometric redshift closer to the actual, spectroscopic redshift
than obtained with M\,82 or Arp\,220.

The smallest average $z_{\rm phot}-z_{\rm spec}$ and standard 
deviation is obtained with 
the II\,Zw\,40 model having $A_V=20$\,mag and a thermal radio spectrum.
If we were to calculate photometric redshifts of the objects in this
sample using this SED, 
we would obtain a mean and standard deviation of 
$\langle z_{\rm phot}-z_{\rm spec}\rangle\,=\,0.02\,\pm\,0.50$.
If we use \sbs\ ($A_V=30$\,mag $\alpha=-0.3$),
we would obtain a mean and standard deviation of 
$\langle z_{\rm phot}-z_{\rm spec}\rangle\,=\,0.17\,\pm\,0.61$.
Were we to use Arp\,220 (M\,82), we would obtain results 
inferior to both of these
with large means and significantly larger scatters:
$\langle z_{\rm phot}-z_{\rm spec}\rangle\,=\,0.8 (2.5)\,\pm\,1.0 (1.5)$.

The reason that BCD SEDs better match
the observed SEDs in these high-z submm galaxies with
Spitzer IRAC$+$MIPS data is a combination of some or all
of the following effects: 1) the $\la5\mu$m region (stellar light)
is much more absorbed in some BCDs than in the M\,82 and Arp\,220 templates;
2) all \citep[active, see][]{hunt05} BCDs are featureless in the mid-IR; 
3) the infrared bump is broader in some BCD SEDs; 
4) the flatter thermal radio continuum of some BCDs better
matches the radio observations in some sources.


\section{BCDs versus Submm Galaxies with only Radio Data \label{sec:radio}}

Thanks to the recent spectroscopic
surveys of submm sources with radio counterparts, large samples
of submm sources with secure spectroscopic redshifts are now
available \citep[e.g.][]{chapman05}. For most
of these sources Spitzer data are not yet published, making unfeasible
a detailed analysis such as the one performed in the previous section. 
However, we can test the photometric redshift technique
by exploiting the 850$\mu$m and radio data, as was done for most
submm sources in the past.

We have compiled a list of all SCUBA sources
with spectroscopic redshifts \citep{chapman02,chapman05}, discarding those lensed and with multiple
counterparts to avoid complications that may arise in the interpretation
of the SED in these cases. We have also discarded sources whose spectra
show clear signatures of an AGN, since in this Letter we
focus on the interpretation of the SEDs due to star formation alone.
These selection criteria resulted in a sample of 40 sources, which are
listed in Table \ref{tab:radio}. 

\begin{figure}
\includegraphics[angle=0,width=0.50\linewidth,bb=160 318 370 670]{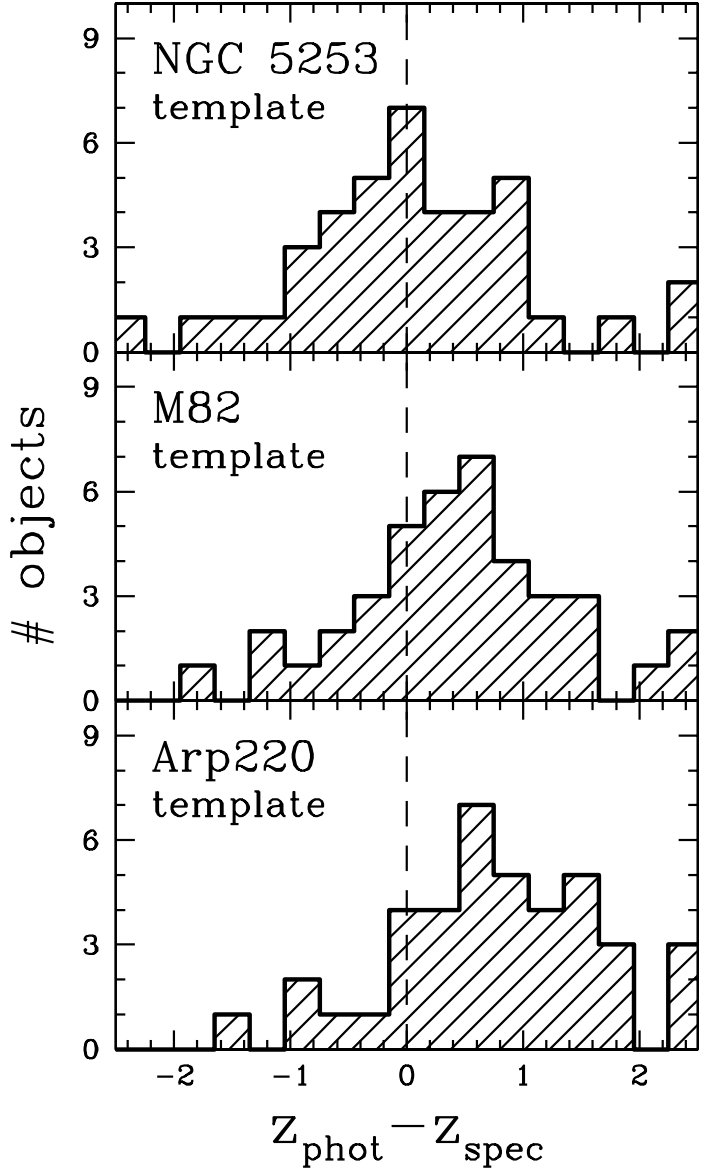}
\caption{Histograms of the $\rm z_{phot}-z_{spec}$ best-fit BCD template, NGC\,5253,
and the two ``standard'' templates, M\,82 and Arp\,220.
The spread is similar for all three templates, but the zero mean
of NGC\,5253 is evident.
\label{fig:his}}
\end{figure}

For each source we estimated the submm-to-radio photometric redshift
by using both the BCD templates and the ``classical' templates M\,82
and Arp\,220, as discussed in the previous section. We found that in 
63\% (25/40) of the cases, BCD templates provide a photometric redshift
that is closer to the spectroscopic one than with M\,82 or Arp\,220.
Moreover, the mean difference $\langle \rm z_{phot}-z_{spec}\rangle$ averaged over the
sample is significantly reduced for the BCD templates ($\sim0.1$ for NGC\,5253), 
relative to M\,82 ($\sim0.5$) or Arp220 ($\sim0.8$).
If we analyze all 62 objects, including those with known AGN/QSO emission,
we obtain a similar result: in
61\% (38/62) of the cases, BCD templates provide a better photometric redshift
than either M\,82 or Arp\,220.
These results are summarized in Table \ref{tab:radio}, while Fig. \ref{fig:his} 
shows the distribution of $\rm z_{phot}-z_{spec}$ for different templates.
It is clear that the BCD template for NGC\,5253 gives
a distribution much more peaked around zero.

For this sample of 40 objects, whose SEDs comprise only submm and radio fluxes,
the SED which best predicts $\rm z_{phot}$ is NGC\,5253, 
with an $A_V=10$\,mag and a thermal radio spectrum.
More than 60\% of the radio-identified submm sources are better fit with
a flatter radio spectrum ($-0.1\le\alpha\le-0.3$)
than with the usual steeper spectrum
$\alpha\approx-0.7$, observed in virtually all evolved starbursts \citep{condon92}. 

\section{Conclusions}

``Active'' BCDs may be characterized by star formation
modes similar to those in high-z starburst galaxies,
specifically for the
efficiency and compactness of the star formation.
We have compared the observed SEDs of spectroscopically identified
submm galaxies with BCD SED templates, for 
both a small sample of submm galaxies with sensitive Spitzer
observations, and a larger sample of submm galaxies with only radio
(and submm) data. In most cases we find that BCD galaxies
match the SED observed in distant submm galaxies better than the
more widely adopted templates of ``standard'' starbursts (M\,82 and Arp\,220).
If the BCD templates are used for the photometric redshift technique,
then the inferred $\rm z_{phot}$ are significantly closer to the actual
spectroscopic redshifts than what is inferred by using the templates
of M\,82 or Arp\,220.
Our results strongly suggest that the the dusty embedded
star clusters in BCD galaxies provide
superior templates to derive photometric redshift of distant submm,
starburst galaxies which may be too faint to be identified spectroscopically.

\acknowledgments

We are grateful to Simone Bianchi,
Reinhard Genzel, Linda Tacconi for insight and critical comments.

\clearpage

\clearpage

\clearpage

\begin{deluxetable}{lrccccccccc}
\tabletypesize{\scriptsize}
\tablecaption{Photometric redshifts $\rm z_{phot}-z_{spec}$ and residuals for 
submm-selected SCUBA sources with VLA and Spitzer data\label{tab:zphot}}
\tablewidth{0pt}
\tablehead{
&&&\multicolumn{4}{c}{II\,Zw\,40} 
&\multicolumn{1}{c}{NGC\,5253} 
&\multicolumn{1}{c}{\sbs} 
&\multicolumn{1}{c}{M\,82} 
&\multicolumn{1}{c}{Arp\,220} \\
\cline{4-7} \\
&&&\multicolumn{2}{c}{$\alpha=-0.5$} 
&\multicolumn{2}{c}{$\alpha=-0.1$} 
&\multicolumn{1}{c}{$\alpha=-0.1$} 
&\multicolumn{1}{c}{$\alpha=-0.3$} \\
\cline{4-5}
\cline{6-7} \\
&&&\multicolumn{1}{c}{$A_V=30$} 
&\multicolumn{1}{c}{$A_V=20$} 
&\multicolumn{1}{c}{$A_V=30$} 
&\multicolumn{1}{c}{$A_V=20$} 
&\multicolumn{1}{c}{$A_V=10$} 
&\multicolumn{1}{c}{$A_V=30$} \\
\colhead{Source}&\colhead{z$_{\rm spec}$\tablenotemark{a}} &\colhead{$\langle$Res.$\rangle$\tablenotemark{b}}
&\colhead{(1)} &\colhead{(2)} 
&\colhead{(3)} &\colhead{(4)} 
&\colhead{(5)} &\colhead{(6)} 
&\colhead{(7)} &\colhead{(8)}
}
\startdata
CFRS14.1157   &  1.15  & 0.15 (5) & -0.75     & -0.75     & -0.65     & -0.65     & -0.35     &  0.95     & -0.35     & -0.95     \\
Frayer.199    &  1.06  & 0.32 (8) & -0.46    &  0.64     & -0.36     &  0.64     &  1.64     &  0.34     &  2.14     &  1.64     \\
LE850.1       &  2.15 & 0.36 (2) & -0.95     &  0.45     & -0.95     &  0.45     &  1.65     &  0.25     &  3.85     &  1.05     \\
LE850.8       &  0.85 & 0.44 (4) & -0.35     &  0.35     & -0.35     &  0.25     &  0.85     &  0.55     &  2.35     &  1.85     \\
LE850.14      &  2.40  & 0.23 (4) & -1.40     & -0.30     & -1.40     & -0.30     &  0.80     & -0.40     &  3.40     &  0.80     \\
LE850.18      &  2.69 & 0.25 (8) & -1.69     & -0.29     & -1.69     & -0.29     &  0.81     & -0.69     &  3.31     &  0.51     \\
\\
$\langle dz \rangle$ &&           & -0.93     &  0.02     & -0.90     &  0.02     &  0.90     &  0.17     &  2.45     &  0.82     \\
Std.dev.             &&           &  0.53     &  0.54     &  0.55     &  0.50     &  0.73     &  0.61     &  1.52     &  1.00     \\
\enddata
\tablenotetext{a}{z$_{spec}$ from \citet{higdon04}; \citet{frayer04}; 
\citet{chapman05}; \citet{ivison04}; \citet{chapman05}; \citet{egami04}.}
\tablenotetext{b}{\,Lowest residuals for z=z$_{\rm spec}$, obtained
with the template in parentheses.}
\end{deluxetable}    

\clearpage

\begin{deluxetable}{lrcrrrrrrrr}
\tabletypesize{\scriptsize}
\tablecaption{Photometric redshifts $\rm z_{phot}-z_{spec}$ and residuals for 
submm-selected SCUBA sources with VLA data only \label{tab:radio}}
\tablewidth{0pt}
\tablehead{
&&&\multicolumn{4}{c}{II\,Zw\,40} 
&\multicolumn{1}{c}{NGC\,5253} 
&\multicolumn{1}{c}{\sbs} 
&\multicolumn{1}{c}{M\,82} 
&\multicolumn{1}{c}{Arp\,220} \\
\cline{4-7} \\
&&&\multicolumn{2}{c}{$\alpha=-0.5$} 
&\multicolumn{2}{c}{$\alpha=-0.1$} 
&\multicolumn{1}{c}{$\alpha=-0.1$} 
&\multicolumn{1}{c}{$\alpha=-0.3$} \\
\cline{4-5}
\cline{6-7} \\
&&&\multicolumn{1}{c}{$A_V=30$} 
&\multicolumn{1}{c}{$A_V=20$} 
&\multicolumn{1}{c}{$A_V=30$} 
&\multicolumn{1}{c}{$A_V=20$} 
&\multicolumn{1}{c}{$A_V=10$} 
&\multicolumn{1}{c}{$A_V=30$} \\
\colhead{Source\tablenotemark{a}}&\colhead{z$_{\rm spec}$\tablenotemark{a}} &\colhead{$\langle$Res.$\rangle$\tablenotemark{b}}
&\colhead{(1)} &\colhead{(2)} 
&\colhead{(3)} &\colhead{(4)} 
&\colhead{(5)} &\colhead{(6)} 
&\colhead{(7)} &\colhead{(8)}
}
\startdata
FN1-40        &  0.45  &   3      &  0.35     &  0.45     &  0.25     &  0.35     &  0.35     &  4.45     &  0.65     &  1.05     \\
FN1-64        &  0.91  &   1      & -0.01     &  0.09     & -0.11     & -0.01     &  0.09     &  4.39     &  0.29     &  0.59     \\
030231.81+001031.3  &  1.32  &   3      &  1.28     &  1.38     &  1.18     &  1.28     &  1.18     &  5.68     &  1.58     &  1.88     \\
030236.15+000817.1  &  2.44  &   7      & -0.23     & -0.14     & -0.34     & -0.23     & -0.34     &  4.57     &  0.07     &  0.47     \\
030244.82+000632.3  &  0.18  &   3      &  1.22     &  1.32     &  1.02     &  1.12     &  1.02     &  6.82     &  1.42     &  1.82     \\
105151.69+572636.0  &  1.15  &   3      &  0.65     &  0.75     &  0.55     &  0.55     &  0.55     &  5.85     &  0.95     &  1.35     \\
\\
$\langle dz \rangle$                 &&           &  0.25     &  0.26     &  0.14     &  0.19     &  0.10     &  4.89     &  0.47     &  0.79     \\
Std.dev.             &&           &  1.13     &  1.05     &  1.20     &  1.18     &  1.12     &  0.83     &  1.05     &  1.02     \\

\enddata
\tablenotetext{a}{Sources and z$_{\rm spec}$ from \citet{chapman02,chapman05}.}
\tablenotetext{b}{\,Lowest residuals for z=z$_{\rm spec}$ obtained
with this template.}
[The complete version of this table is in the electronic edition of
the Journal.  The printed edition contains only a sample.]
\end{deluxetable}

\end{document}